\documentclass[letterpaper, 10 pt, conference]{ieeeconf}  
\usepackage{graphicx}      
\usepackage{subcaption}
\usepackage{caption}
\usepackage{amsmath}
\usepackage{amssymb}
\usepackage{algorithm}
\usepackage{algpseudocode}
\usepackage{hyperref}
\usepackage{xcolor}

\IEEEoverridecommandlockouts                              

\title{\LARGE \bf
S-CycleGAN: Semantic Segmentation Enhanced CT-Ultrasound Image-to-Image Translation for Robotic Ultrasonography
}

\author{Yuhan Song and Nak Young Chong
\thanks{This work was supported by JSPS KAKENHI Grant Number JP23K03756.}
\thanks{Both authors are with the School of Information Science, Japan Advanced Institute of Science and Technology, Nomi, Ishikawa 923-1292, Japan {\tt\small \{yuhan-s, nakyoung\}@jaist.ac.jp}}
}

\begin{document}

\maketitle
\thispagestyle{empty}
\pagestyle{empty}

\begin{abstract}

Ultrasound imaging is pivotal in various medical diagnoses due to its non-invasive nature and safety. In clinical practice, the accuracy and precision of ultrasound image analysis are critical. Recent advancements in deep learning are showing great capacity of processing medical images. However, the data hungry nature of deep learning and the shortage of high-quality ultrasound image training data suppress the development of deep learning based ultrasound analysis methods. To address these challenges, we introduce an advanced deep learning model, dubbed S-CycleGAN, which generates high-quality synthetic ultrasound images from computed tomography (CT) data. This model incorporates semantic discriminators within a CycleGAN framework to ensure that critical anatomical details are preserved during the style transfer process. The synthetic images are utilized to enhance various aspects of our development of the robot-assisted ultrasound scanning system. The data and code will be available at \href{https://github.com/yhsong98/ct-us-i2i-translation}{\color{magenta}{https://github.com/yhsong98/ct-us-i2i-translation}}.

\end{abstract}

\section{INTRODUCTION}

Ultrasound imaging is one of the most widely implemented medical imaging modalities, offering a versatile, non-invasive, and cost-effective method for visualizing the internal structures of the body in real-time. Although ultrasound imaging is safe and convenient, analyzing these images presents considerable challenges due to factors such as low contrast, acoustic shadows, and speckles \cite{8614204}. Deep learning based medical image processing methods have made great breakthroughs in recent years and have been the state-of-the-art tool for medical image processing applications in various fields, including detection, segmentation, classification, and synthesis \cite{zhou2023deep}. 

Nonetheless, due to the data-hungry nature of deep learning, the performance of those methods relies heavily on a large amount of image data and manual annotations. While progress in unsupervised learning techniques and the emergence of large-scale open-source image datasets have mitigated these issues somewhat, these solutions are less applicable in the field of medical image processing due to several factors \cite{li2023systematic}. Firstly, medical images require precise and reliable annotations, which must often be provided by expert clinicians, making the process time-consuming and expensive. Secondly, patient privacy concerns limit the availability and sharing of medical datasets. Moreover, the variability in medical imaging equipment and protocols across different healthcare facilities can lead to inconsistencies in the data, complicating the development of generalized models. 

\begin{figure}[h!]
	\centering
	\includegraphics[width=\linewidth]{"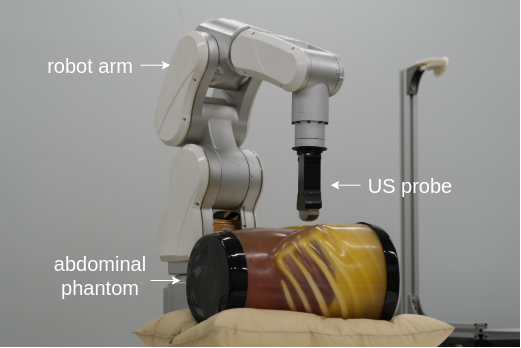"}
	\caption{Our robot-assisted ultrasound imaging system}
	\label{fig:russ}
\end{figure}

Along the lines, we are building a fully automated robot-assisted ultrasound scan system (RUSS). This platform is designed to perform abdominal ultrasound scans without any human intervention (Fig.~\ref{fig:russ}). Thus we have proposed several versions of ultrasound image segmentation algorithms as evaluation metrics for the robot arm movements \cite{song2023abdominal,song2023two,song2024abdominal}. However, our prior efforts have been restricted by limited data sources. While our segmentation algorithms have demonstrated effectiveness within our experimental datasets, we anticipate that training our model with a more diverse array of data would enhance its robustness and general applicability. Furthermore, we aim to create a simulation environment to facilitate the development of our RUSS, allowing for refined testing and optimization under controlled conditions. A pre-operative 3D model reconstructed from CT scans are planned to be utilized as the scan target. Based on the current contact point and angle of the virtual ultrasound probe, the system will generate and provide a corresponding ultrasound image as feedback. This integration will enable the RUSS to simulate realistic scanning scenarios, allowing for precise alignment and positioning adjustments that reflect actual clinical procedures.

In this research, we proposed a semantically enhanced CycleGAN, dubbed S-CycleGAN. By adding additional segmentation models as semantic discriminators, together with the original style discriminator, the proposed model is capable of transferring the style of CT slice to the ultrasound domain while keeping the transformed image semantically consistent with the source image.

\section{RELATED WORK}

\subsection{Image-to-image translation}\label{sec:related_work_i2i}

The process of image-toimage translation is fundamental in various applications, ranging from artistic style transfer to synthesizing realistic datasets. One seminal work in this field is the introduction of the Generative Adversarial Network (GAN) by Goodfellow et al. \cite{goodfellow2014generative}. The GAN framework involves a dual-network architecture where a generator network competes against a discriminator network, fostering the generation of highly realistic images. Building on this, Zhu et al. introduced CycleGAN \cite{zhu2017unpaired}, which allows for image-to-image translation in the absence of paired examples. In the context of medical imaging, Sun et al. \cite{sun2023double} leveraged a double U-Net CycleGAN to enhance the synthesis of CT images from MRI images. Their model incorporates a U-Net-based discriminator that improves the local and global accuracy of synthesized images. Chen et al. \cite{CHEN2024106100} introduced a correction network module based on an encoder-decoder structure into a CycleGAN model. Their module incorporates residual connections to efficiently extract latent feature representations from medical images and optimize them to generate higher-quality images.

\subsection{Ultrasound image synthesis}

As for medical ultrasound image synthesis, advancements have been achieved due to the integration of deep learning techniques, particularly GANs and Denoising Diffusion Probabilistic Models (DDPMs) \cite{ho2020denoising}. Liang et al. \cite{liang2022sketch} employed GANs to generate high-resolution ultrasound images from low-resolution inputs, thereby enhancing image clarity and detail that are crucial for effective medical analysis. Stojanovski et al. \cite{stojanovski2023echo} introduced a novel approach to generating synthetic ultrasound images through DDPM. Their study leverages cardiac semantic label maps to guide the synthesis process, producing realistic ultrasound images that can serve as substitutes for actual data in training deep learning models for tasks like cardiac segmentation.

In the specific context of synthesizing ultrasound images from CT images, Vitale et al. \cite{vitale2020improving} proposed a two-stage pipeline. Their method begins with the generation of intermediate synthetic ultrasound images from abdominal CT scans using a ray-casting approach. Then a CycleGAN framework operates by training on unpaired sets of synthetic and real ultrasound images. Song et al. \cite{song2022ct2us} also proposed a CycleGAN based method to synthesize ultrasound images from abundant CT data. Their approach leverages the rich annotations of CT images to enhance the segmentation network learning process. The segmentation networks are initially pretrained on the synthetic dataset translated from preprocessed CT images. Then they are fine-tuned on actual ultrasound images to refine their ability to accurately segment kidneys.

\section{METHODOLOGY}

\subsection{CycleGAN}

\begin{figure*}[ht]
	\centering
	\includegraphics[width=0.9\linewidth]{"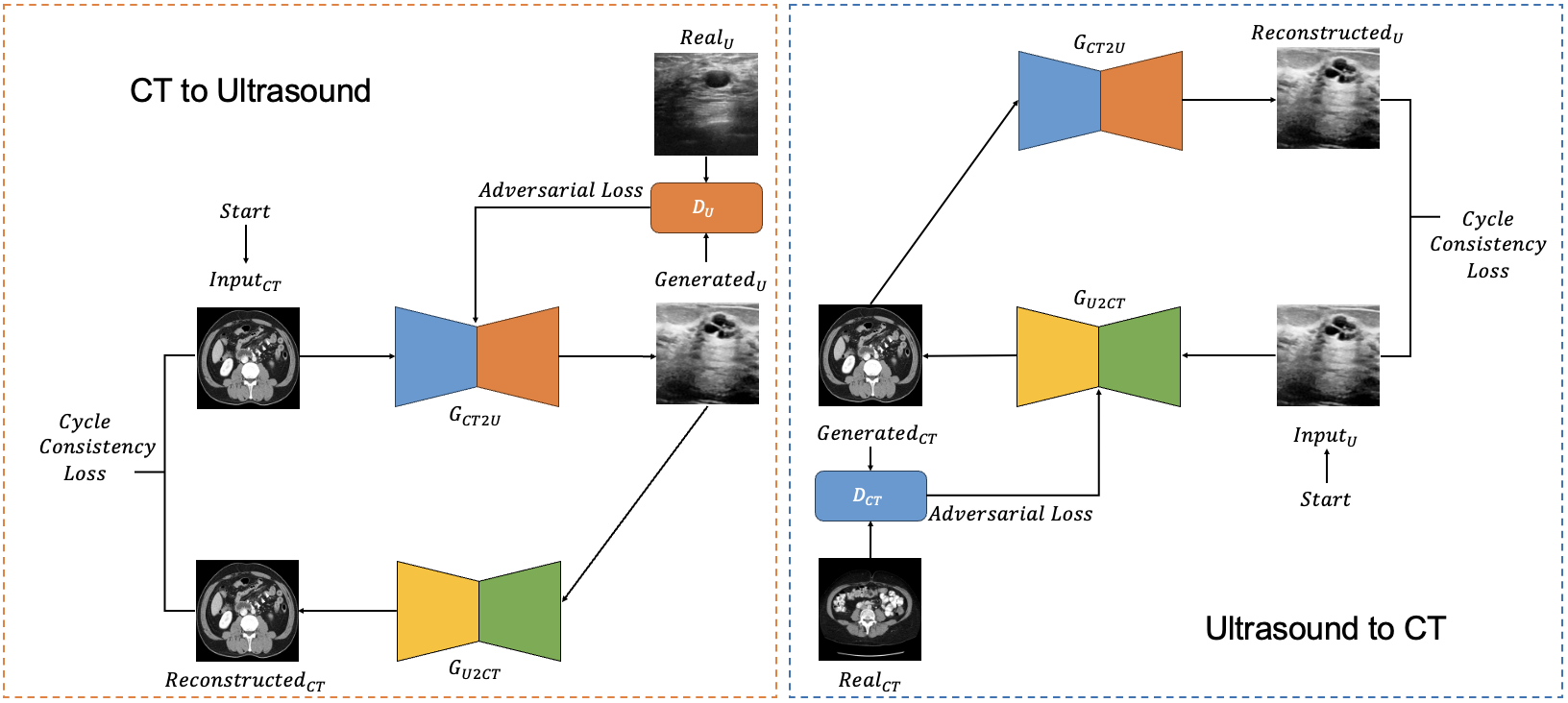"}
	\caption{CycleGAN for CT-ultrasound translation}
	\label{fig:cyclegan}
\end{figure*}

In this research, we aim to translate CT images into ultrasound images. Conventionally, one might consider training a neural network that inputs a CT image and outputs its corresponding ultrasound image, followed by computing the similarity between the synthesized ultrasound image and the actual ultrasound image to update the network. However, we face a challenge: our datasets, one containing abdominal CT volumes \cite{Ma-2021-AbdomenCT-1K} and the other comprising abdominal ultrasound images \cite{vitale2020improving}, are unpaired. Given this situation, we have chosen to employ CycleGAN, as it is designed for image-to-image translation tasks where paired images are unavailable. The architecture of CycleGAN includes four key components: two generator networks and two discriminator networks. The generators are responsible for translating images from one domain (e.g., CT) to another (e.g., ultrasound) and vice versa. Each generator has a corresponding discriminator that aims to distinguish between real images from the target domain and fake images created by the generator. A distinctive feature of CycleGAN is the incorporation of a cycle consistency loss. This design is based on the assumption that for instance, translating a sentence from English to French and then back to English should ideally return the original sentence, they apply a similar concept in the image-to-image translation. Mathematically, if \( G: X \rightarrow Y \) represents a translator from domain \( X \) to domain \( Y \), and \( F: Y \rightarrow X \) serves as its counterpart, then \( G \) and \( F \) should function as inverses to each other, with both mappings being bijections. To enforce this structure, they train the mappings \( G \) and \( F \) concurrently while incorporating a cycle consistency loss \cite{zhou2016learning}. This loss ensures that \( F(G(x)) \approx x \) and \( G(F(y)) \approx y \), promoting fidelity in the translation process between the two domains. If we directly apply the CycleGAN to our task, it should follow the pipeline in Fig.~\ref{fig:cyclegan}.

\subsection{Proposed semantic segmentation enhanced S-CycleGAN}
After training and testing an original CycleGAN model, we observed that while the overall style (color and texture) of the CT images was effectively transformed to match the ultrasound style, the anatomical details in the generated ultrasound images are hard to distinguish. This difficulty stems from the fact that in traditional image translation tasks, images from both domains are treated as samples from the joint distribution of all relevant sub-classes (such as different organs), and the translation is essentially a mapping between these distributions. Even with the use of cyclical mappings, there is no assurance that the marginal distributions of these sub-classes (or modes) are properly matched (e.g., `liver' correctly translating to `liver').

\begin{figure}[h!]
        \centering
	\begin{subfigure}{0.45\linewidth}
	\includegraphics[width=\linewidth]{"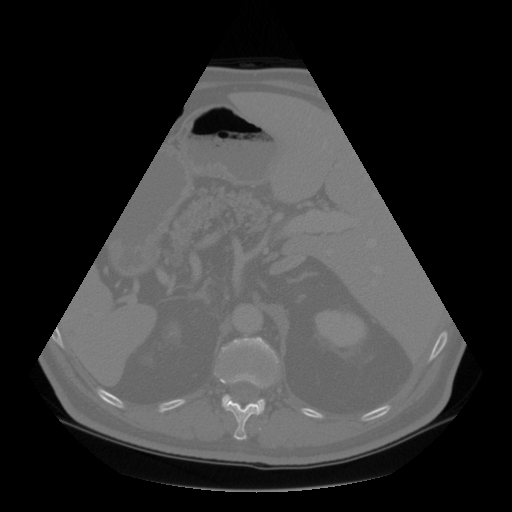"}
	\caption{Real CT}\label{fig:ct_cyclegan}
	\end{subfigure}
	\hfill
	\begin{subfigure}{0.45\linewidth}
	\includegraphics[width=\linewidth]{"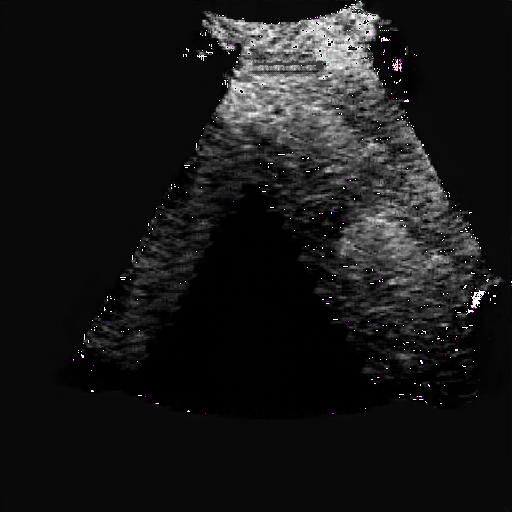"}
	\caption{Fake US}\label{fig:fus_cyclegan}
	\end{subfigure}
	
	\begin{subfigure}{0.45\linewidth}
	\includegraphics[width=\linewidth]{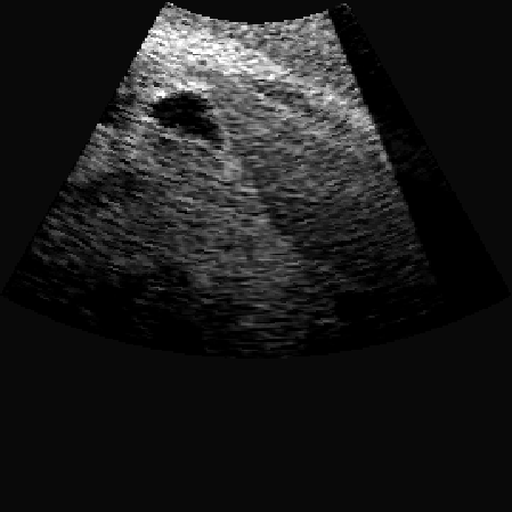}
	\caption{Real US}\label{fig:us_cyclegan}
	\end{subfigure}
	\hfill
	\begin{subfigure}{0.45\linewidth}
	\includegraphics[width=\linewidth]{"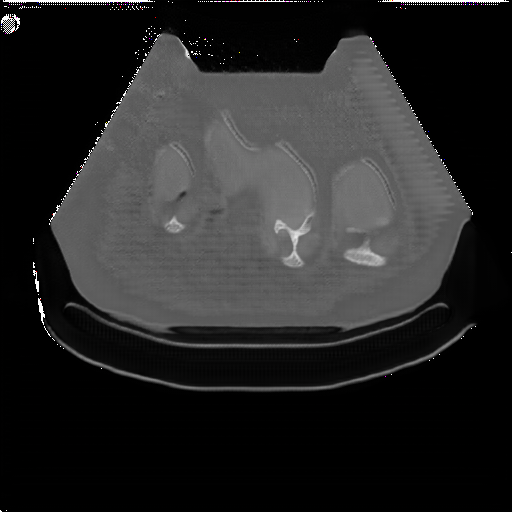"}
	\caption{Fake CT}\label{fig:fct_cyclegan}
	\end{subfigure}

        \begin{subfigure}{0.45\linewidth}
	\includegraphics[width=\linewidth]{"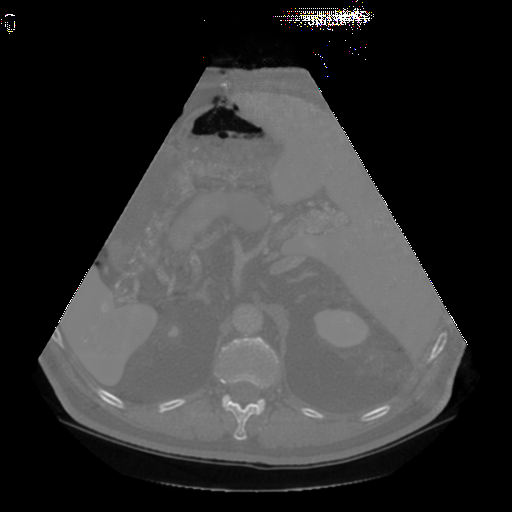"}
	\caption{Reconstructed CT}\label{fig:reca_cyclegan}
	\end{subfigure}
        \hfill
	\begin{subfigure}{0.45\linewidth}
	\includegraphics[width=\linewidth]{"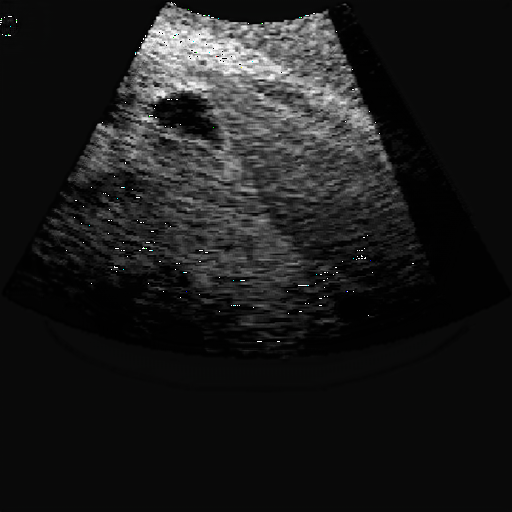"}
	\caption{Reconstructed US}\label{fig:recb_cyclegan}
	\end{subfigure}

	\caption{CT-ultrasound translation using pure CycleGAN}
	\label{fig:cycleganres}
\end{figure}

\begin{figure*}[h]
        \centering
	\begin{subfigure}{0.9\linewidth}
	\includegraphics[width=\linewidth]{"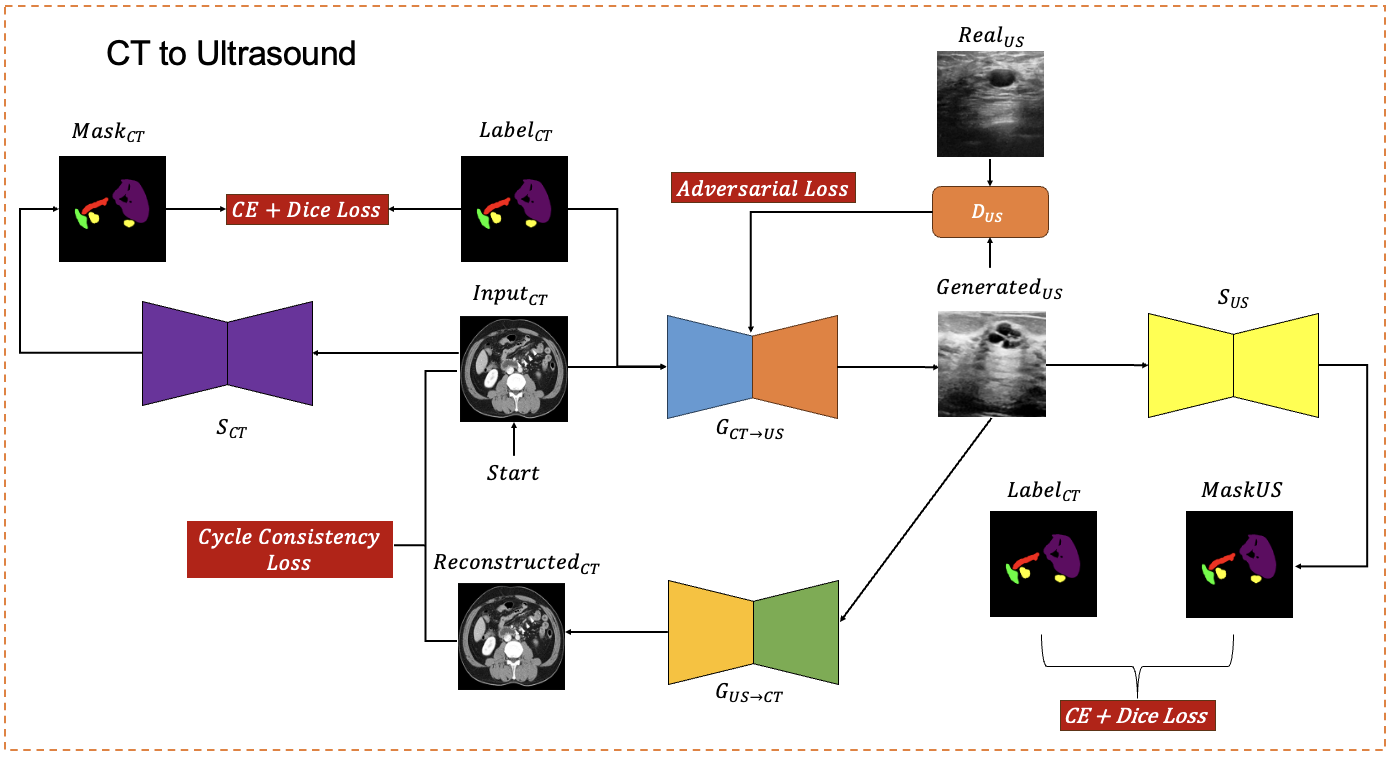"}
	\caption{CT-to-Ultrasound}
	\end{subfigure}
	\hfill
	\begin{subfigure}{0.9\linewidth}
	\includegraphics[width=\linewidth]{"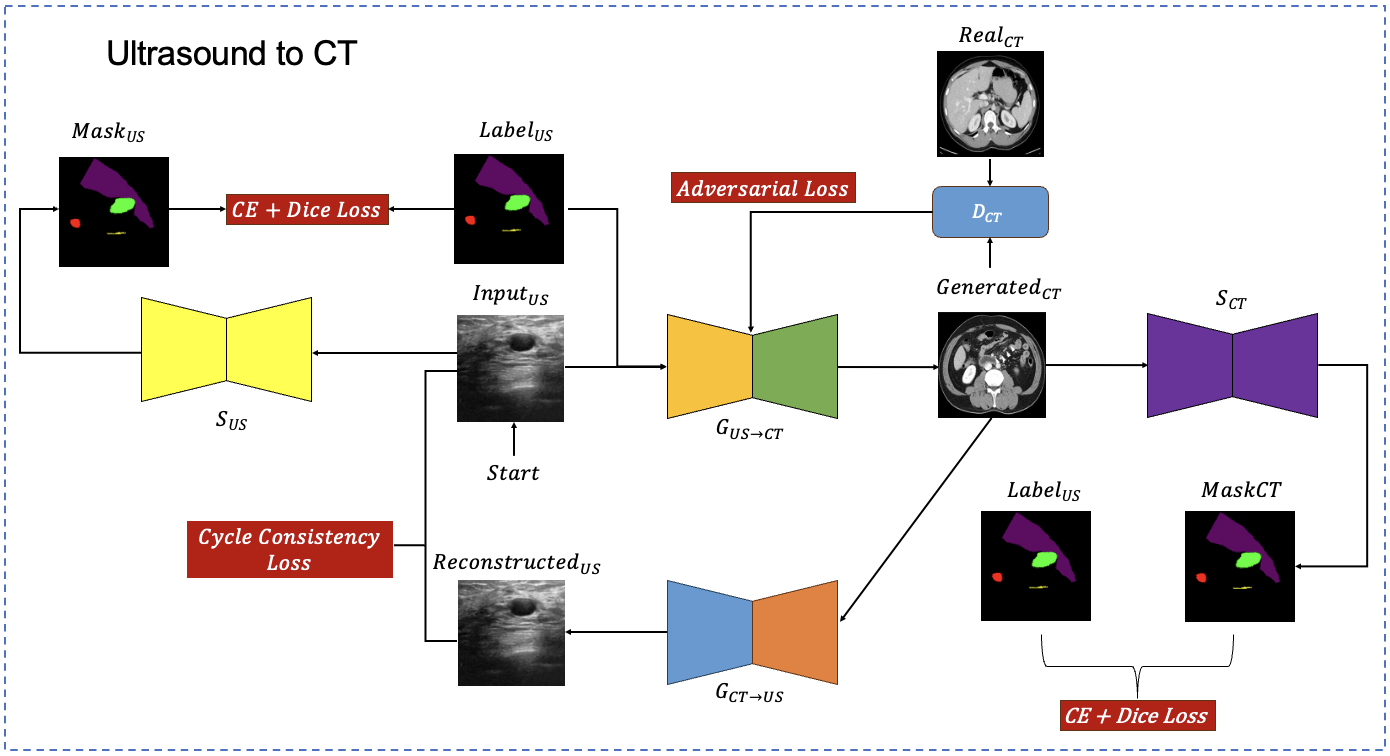"}
	\caption{Ultrasound-to-CT}
	\end{subfigure}
        \caption{Proposed Pipeline}
        \label{fig:s-cyclegan}
\end{figure*}

To maintain pixel-level semantic accuracy while converting image-level style (color and texture distribution), we incorporated two additional segmentation networks as semantic discriminators. The fake images produced by the generator are analyzed by these segmentation networks to produce a semantic mask. Subsequently, a segmentation loss is computed between this semantic mask and the label of the real source image. Moreover, unlike other similar studies \cite{cherian2019sem} that continue to use an image alone as input, our network architecture employs both the image and its corresponding semantic map as inputs. This dual-input approach equips the generator with a more refined understanding of per-pixel semantic information. Hereby, we propose our S-CycleGAN (Fig.~\ref{fig:s-cyclegan}), which includes the following components:

\subsubsection{Generators $G_{CT \to US}$ and $G_{US \to CT}$}
These networks translate images from CT to ultrasound ($G_{CT \to US}$) and ultrasound to CT ($G_{US \to CT}$) respectively. They are trained to minimize both the adversarial and cycle consistency losses to produce realistic translations.

\subsubsection{Discriminators $D_{US}$ and $D_{CT}$}
$D_US$ discriminates between real and generated ultrasound images, whereas $D_{CT}$ differentiates real CT images from those generated by $G_{US \to CT}$. They enforce the adversarial loss component, pushing generators to create indistinguishable images from real ones.

\subsubsection{Segmentation Networks ($S_{US}$ and $S_{CT}$)}
$S_{US}$ and $S_{CT}$ are responsible for generating semantic masks from ultrasound and CT images, respectively, to ensure that critical anatomical features are retained during translation.

\subsubsection{Adversarial Loss}

\begin{equation}
    \begin{aligned}
        \mathcal{L}_{adv}^{CT\to US} = &\mathbb{E}_{us \sim p_{\text{data}}({us})}[\log D_{US}({us})] \\
        + &\mathbb{E}_{ct \sim p_{\text{data}}({ct})}[\log (1 - D_{US}(G_{CT \to US}({ct})))]
    \end{aligned}
\end{equation}


where $G_{CT\to US}$ tries to generate images that look similar to ultrasound images , while $D_{US}$ aims to distinguish between translated samples and real ultrasound image samples. $G_{CT\to US}$ aims
to minimize this objective against an adversary $D_{US}$ that tries to maximize it. The counterpart is for $G_{US\to CT}$ and $D_{CT}$ is vice versa.

\subsubsection{Cycle Consistency Loss}
    \begin{equation}
    \begin{aligned}
        \mathcal{L}_{cycle} = &\| G_{US \to CT}(G_{CT \to US}(x_{CT})) - x_{CT} \|_1 \\
        + &\| G_{CT \to US}(G_{US \to CT}(x_{US})) - x_{US} \|_1
    \end{aligned}
    \end{equation}
Cycle consistency loss ensures that translating an image to the other domain and back again will yield the original image, maintaining cycle consistency across translations, which is the key to training image-to-image translation models with unpaired image sets. The L1 norm here measures the absolute differences between the original and the reconstructed image.

\subsubsection{Segmentation Loss}
The segmentation loss in the CycleGAN architecture is a combination of Cross-Entropy Loss and Dice Loss. And there are two segmentation losses computed separately on each domain: $\mathcal{L}_{seg}^{CT}$, and $\mathcal{L}_{seg}^{US}$.

\begin{equation}
    \mathcal{L}_{CE} = -\sum_{i=1}^C y_{i} \log(p_{i})
\end{equation}

\begin{equation}
    \mathcal{L}_{Dice} = 1 - \sum_{i=1}^{C} \frac{2 y_i \cdot p_i}{y_i + p_i +\epsilon}
\end{equation}
where $y_i$ and $p_i$ are the groundtruth and predicted probability of being of class $i$. And $C$ is the number of classes. $\epsilon$ is an arbitrarily small smooth parameter. 

The propagation flow is provided in Algorithm~\ref{al:cyclegan}.
\begin{algorithm}[h]
\caption{S-CycleGAN Training}
\label{al:cyclegan}
\begin{algorithmic}[1]
\State \textbf{Input:} Source domain $CT$, target domain $US$, hyperparameters $\lambda_{cycle}$, and $\lambda_{seg}$
\State Initialize networks $G_{CT \to US}$, $G_{US\to CT}$ for generation, $D_{CT}$, $D_{US}$ for discrimination, $S_{CT}$, $S_{US}$ for semantic segmentation.

\For{each epoch}
    \For{each batch}
        \State \textit{Forward Pass:}
        \State $real\_CT, real\_CT\_Mask \gets$ sample from $CT$
        \State $real\_US, real\_US\_Mask \gets$ sample from $US$
        \State $fake\_US \gets G_{CT\to US}(real\_CT)$
        \State $rec\_CT \gets G_{US\to CT}(fake\_US)$
        \State $fake\_CT \gets G_{US\to CT}(real\_US)$
        \State $rec\_US \gets G_{CT\to US}(fake\_CT)$
        
        \State \textit{Backward Pass and Optimization:}
        \State Freeze $S_{CT}$, $S_{US}$, $D_{CT}$, $D_{US}$
        \State Update $G_{CT\to US}$, $G_{US\to CT}$ using $\mathcal{L}_{adv}^{CT\to US}+\mathcal{L}_{adv}^{US\to CT}+\lambda_{cycle}\mathcal{L}_{cycle}+\lambda_{seg}(\mathcal{L}_{seg}^{CT}+\mathcal{L}_{seg}^{US})$
        \State Unfreeze $D_{US}$, $D_{CT}$ 
        \State Update $D_{US}$, $D_{CT}$ using $\mathcal{L}_{adv}^{CT\to US}$, $\mathcal{L}_{adv}^{US\to CT}$, respectively 
        \State Freeze $G_{CT\to US}$, $G_{US\to CT}$, $D_{US}$, $D_{CT}$
        \State Unfreeze $S_{CT}$, $S_{US}$
        \State Update semantic segmentors $S_{CT}$, $S_{US}$ using $\mathcal{L}_{seg}^{CT}$, $\mathcal{L}_{seg}^{US}$, respectively
    \EndFor
\EndFor
\end{algorithmic}
\end{algorithm}

\section{EXPERIMENTS and RESULTS}

\subsection{Dataset}
In this study, the CT data is sourced from the AbdomenCT-1K dataset \cite{Ma-2021-AbdomenCT-1K}, while the ultrasound data is obtained from the Kaggle US simulation \& segmentation dataset \cite{vitale2020improving}. Both datasets contain scans from the abdominal region. The CT dataset is annotated with four anatomical structures: liver, kidney, spleen, and pancreas. Conversely, the ultrasound dataset includes annotations for eight anatomical structures: liver, kidney, spleen, pancreas, vessels, adrenals, gallbladder, and bones. Therefore, for this research, we focus on the overlapping structures between the two datasets as the anatomical structures of interest. The specific organs and their corresponding mask colors are detailed in Table~\ref{tb:label}.

\begin{table}[H]
    \centering
    \caption{Organ name and mask color}
    \begin{tabular}{|c|c|c|c|c|}
    \hline
       Organ & Liver & Kidney & Spleen & Pancreas\\ \hline
       Color & Violet & Yellow & Pink & Blue\\
    \hline
    \end{tabular}   
    \label{tb:label}
\end{table}

The Abdomen-1K dataset provides more than 1000 CT scans, and the data is provided in 3D format. Firstly, we randomly select 200 CT scans, and for each CT scan, we randomly sampled 10 transverse plane slices. For a more uniform image shape, we applied a fan shape mask to the CT images to mimic the outline of convex ultrasound images.

\subsection{Network Implementation and Training}

 In our design, the discriminators ($D_{US}$, $D_{CT}$) follow the original design in \cite{zhu2017unpaired}. The generators and segmentation networks are using U-Nets \cite{ronneberger2015u}. The generators take the concatenated image and semantic mask as input, and output a transferred image. The segmentation networks take CT or ultrasound image as input, and output predicted semantic mask. The number of convolutional filters in each U-Net block is 64, 128, and 256. The bottleneck layer has 512 convolutional filters. Initial learning rate is set as 0.0002, and will be decayed after 100 epoch with Adam. Batch size is set to 1, since we only have one GPU. The network is trained with 300 epochs in total. The coefficients $\lambda_{cycle}$, and $\lambda_{seg}$ were experimentally fixed to 10 and 0.5. One thing worth noting is that for the input of the generators and we use RGB images of CT and ultrasound (3 channels). That is for the convenience of generalizing our model to the universal tasks. The segmentation networks are trained before being incorporated into the S-CycleGAN, and using such pre-trained segmentation models help the overall network converge faster. The code implementation is using Pytorch 1.10.1, and we use one RTX 3090 Ti GPU with NVIDIA driver version 550.54.15 and CUDA version 12.4. 

\subsection{Qualitative Results}

\begin{figure}[h]
        \centering
	\begin{subfigure}{0.45\linewidth}
	\includegraphics[width=\linewidth]{"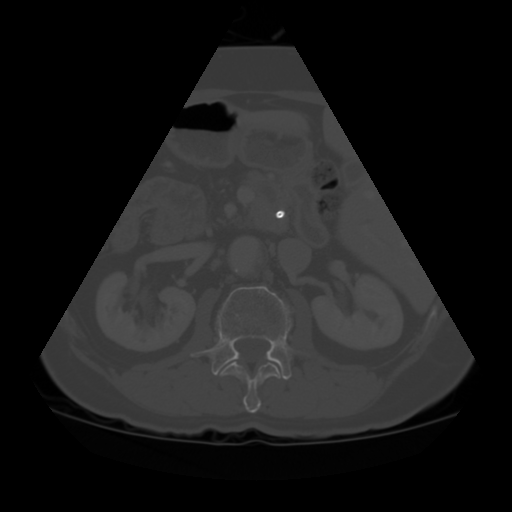"}
	\caption{Real CT}
	\end{subfigure}
	\hfill
	\begin{subfigure}{0.45\linewidth}
	\includegraphics[width=\linewidth]{"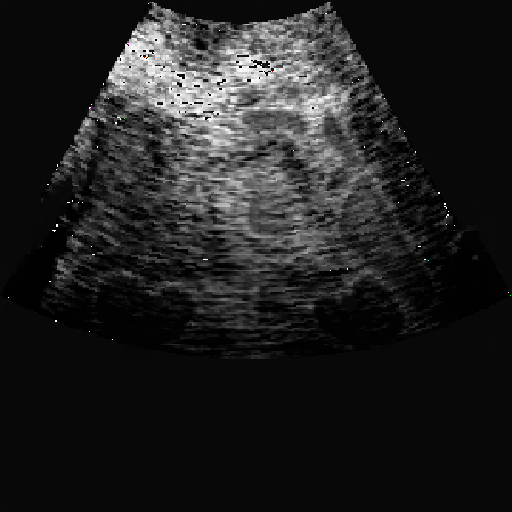"}
	\caption{Fake US}
	\end{subfigure}
   
        \begin{subfigure}{0.45\linewidth}
	\includegraphics[width=\linewidth]{"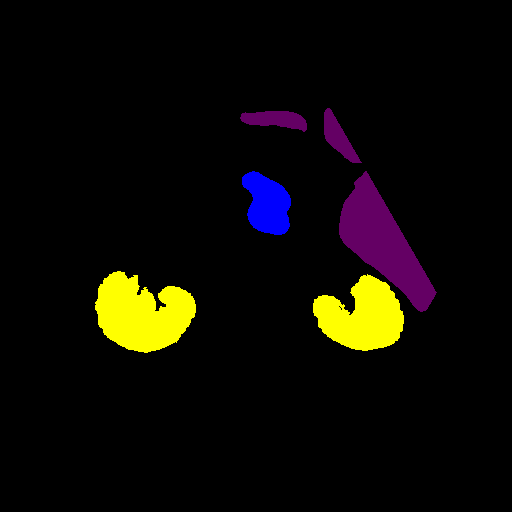"}
	\caption{CT Label}
	\end{subfigure}
	\hfill
	\begin{subfigure}{0.45\linewidth}
	\includegraphics[width=\linewidth]{"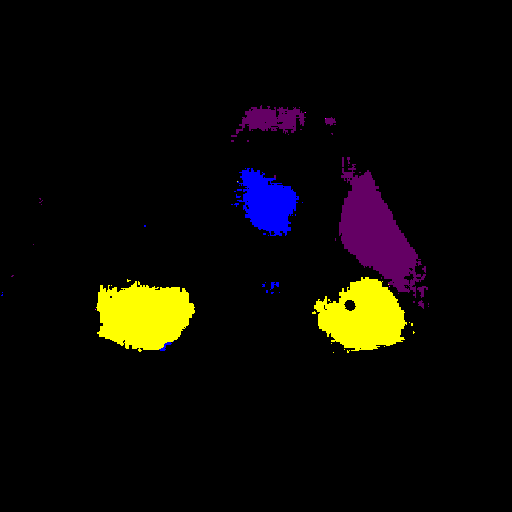"}
	\caption{Predicted US Mask}
	\end{subfigure}
        \caption{CT-to-ultrasound translation example.1}
        \label{fig:finalres_ct2us1}       
\end{figure}

\begin{figure}[h]
        \centering
	\begin{subfigure}{0.45\linewidth}
	\includegraphics[width=\linewidth]{"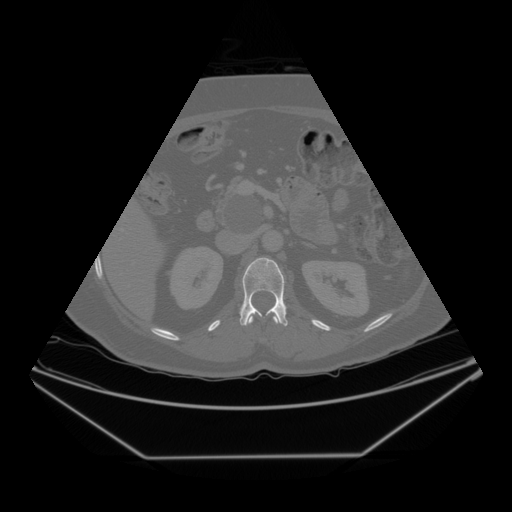"}
	\caption{Real CT}
	\end{subfigure}
	\hfill
	\begin{subfigure}{0.45\linewidth}
	\includegraphics[width=\linewidth]{"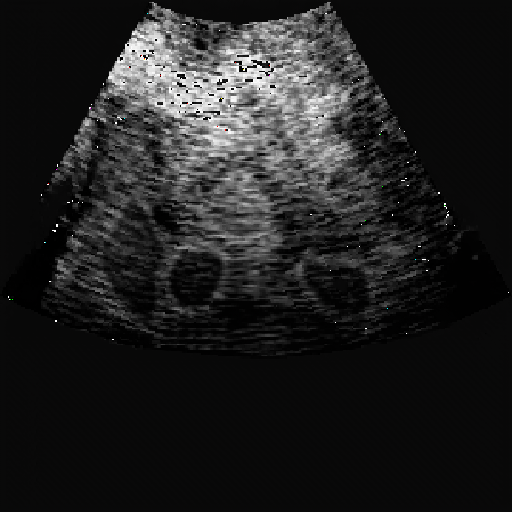"}
	\caption{Fake US}
	\end{subfigure}
   
        \begin{subfigure}{0.45\linewidth}
	\includegraphics[width=\linewidth]{"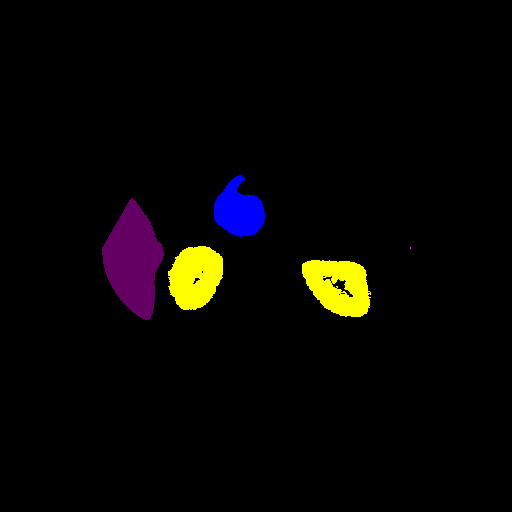"}
	\caption{CT Label}
	\end{subfigure}
	\hfill
	\begin{subfigure}{0.45\linewidth}
	\includegraphics[width=\linewidth]{"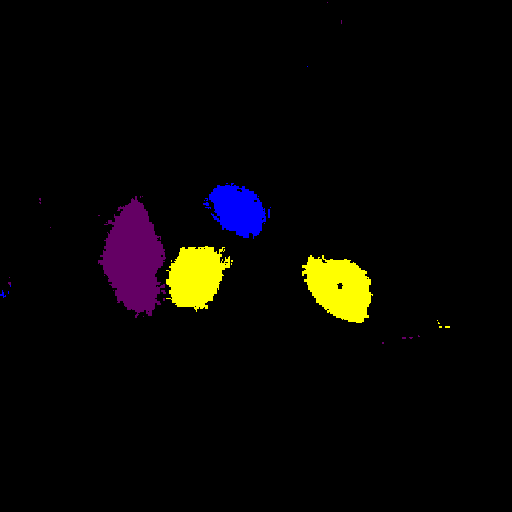"}
	\caption{Predicted US Mask}
	\end{subfigure}
        \caption{CT-to-ultrasound translation example.2}
        \label{fig:finalres_ct2us2}       
\end{figure}

Fig.~\ref{fig:finalres_ct2us1} and \ref{fig:finalres_ct2us2}  present examples of the translation results from CT to ultrasound. These visual comparisons demonstrate that the S-CycleGAN can not only mimic the ultrasound style but also preserve critical anatomical features compared with Fig.~\ref{fig:cycleganres}. The synthetic images closely resemble real ultrasound scans in terms of texture and shape, suggesting a high level of detail preservation. 

\section{CONCLUSION}

This study introduced the S-CycleGAN, adaptation of the CycleGAN framework, enhanced with semantic discriminators for generating synthetic ultrasound images from CT data. The primary innovation of this approach lies in its ability to preserve anatomical details during the image translation process. Our model has demonstrated promising results in generating high-quality ultrasound images that closely replicate the characteristics of authentic scans. These outcomes are significant in the context of medical image translation. However, the current study is not without its limitations. Suitable metrics that comprehensively evaluate the effectiveness of ultrasound image synthesis in a numerical manner are still absent. Future work will include developing these metrics and incorporating feedback from medical experts through structured evaluation protocols. We mentioned that the synthetic dataset is expected to enhance our deep learning models. However, in current stage, the improvement by training deep learning models leveraging synthetic data is minimally significant. While our S-CycleGAN successfully replicates the visual characteristics of ultrasound images, there may still be subtle yet critical differences in textural and anatomical details compared to real ultrasound images. These discrepancies can affect the model's learning process, particularly in tasks requiring high precision. We will continue to explore additional adjustments, including the formulation of a deformation field to accurately simulate the transition from CT to ultrasound imaging and a more proper training process to better leverage the synthetic data. These efforts will not only validate the clinical applicability of the synthetic images but also refine the model's performance to meet the stringent requirements of medical diagnostics.


\bibliographystyle{unsrt}
\bibliography{reference}

\end{document}